\documentclass[twocolumn]{aastex631}
\usepackage{float}
\usepackage{graphicx}
\usepackage{amsmath,amssymb}
\usepackage[section]{placeins}
\usepackage{hyperref}

\begin{document}

\title{How does a low surface brightness galaxy form spiral arms?}

\author{Ganesh Narayanan}
\affiliation{ Department of Physics, Indian Institute of Science Education and Research (IISER) Tirupati, Tirupati - 517507, India}

\author{Anagha A. G.}
\affiliation{ Department of Physics, Indian Institute of Science Education and Research (IISER) Tirupati, Tirupati - 517507, India}

\author{Arunima Banerjee}
\affiliation{ Department of Physics, Indian Institute of Science Education and Research (IISER) Tirupati, Tirupati - 517507, India}



\begin{abstract}

The formation and evolution of spiral arms in low surface brightness galaxies (LSBs) are not well-understood. We study the dynamics of spiral arms in two prototypical LSBs, F568-VI and F568-01, using both analytical models and N-body + hydrodynamical simulations. We first consider the disk as a 2-component system of gravitationally-coupled stars and gas in the force field of a \emph{spherical} dark matter halo, subjected to local, non-axisymmetric perturbations. However, no local spirals are formed. We next assume the disk to be a 1-component system of stars in the net gravitational potential of a galaxy with a \emph{spherical} dark matter halo perturbed by a global $m=2$ instability. In this case, the growth time for spiral formation was low, equal to 0.78 and 0.96 Gyrs, respectively, corresponding to a few dynamical times of the galaxies. Finally, we simulate the LSBs using the N-body + hydrodynamical simulation code RAMSES. \emph{Our results show that a quadrupolar field associated with an oblate halo with an axial ratio of 0.7} is necessary to drive a long-lived global spiral in the LSB disks. Further, feedback corresponding to a supernova mass fraction of $\sim$ 0.05 is essential to comply with the observed stellar surface density. The simulated spirals survives for about ten dynamical times and the average pattern speed lies between 10 - 15  $\rm{kms^{-1}{kpc}^{-1}}$. The spiral arm thus formed is therefore a transient global pattern driven by the tidal field of the oblate dark matter halo.

\end{abstract}


\section{Introduction}
\noindent Low Surface Brightness galaxies (LSBs) are galaxies with central B-band surface brightness $\mathrm{\mu_{B}(0) > 22 \; mag \; arcsec^{-2}}$ \citep{ImpeyBothun}. LSBs are gas-rich \citep{Bothun1985, Schombert2001} and dark matter dominated \citep{de_Blok_2001, Banerjee2010}, and are considered to be under-evolved systems due to the low values of their metallicities and star formation rates. However, they have high gas mass fraction when placed alongside the High Surface Brightness galaxies (HSBs) \citep{Bothun1997P}. These are mostly disk galaxies, and may be located either in the void or isolated environment \citep{Rosenbaum2009A}. The spiral arms constitute the most notable characteristic of face-on disk galaxies, though all disk galaxies may not exhibit these non-axisymmetric features. LSBs mostly exhibit weak and sometimes fragmented spiral structures in their stellar disk \citep{Mcgaugh1995}, the formation and evolution of which have not been studied systematically.\\

\noindent From a dynamical perspective, the formation of spiral arms in galactic disks is primarily triggered by disk instabilities. See, for example, \cite{Binney2008}. \cite{toomre1964gravitational} analytically studied the response of a 1-component, self-gravitating and differentially-rotating, infinitesimally thin disk of fluid, to local axi-symmetric perturbations and introduced the Toomre Q parameter such that $\mathrm{Q > 1}$ would indicate a stable disk. The formalism was later generalized to more realistic multi-component fluid disks, with or without a finite disk thickness \citep{Bertin1988, Jog1996, Romeo_2011}. The response of a 1-component, sheared disk to local, non-axisymmetric perturbations under the WKB or tight-winding approximation was pioneered by \cite{Goldreich1965} and \cite{Julian1966}. \cite{Toomre1981} coined the term swing amplification for this phenomenon since the growth is maximum as a wave swings from a leading to a trailing position. \cite{Jog1992} generalized the study to a 2-component galactic disk model with application to the Galaxy. \cite{Lin1964}, on the other hand, introduced the density wave theory to study the global spiral modes in the galactic disk and derived the dispersion relations using the tight winding approximation. \cite{Jalali_2007} studied the normal modes in a stellar disk modeled by the collisionless Boltzmann equation and found that the unstable modes are regulated by dark matter density distribution in the galaxy.  \\

\noindent The disk dynamics of the LSBs is dominated by the dark matter halo at all radii as is evident from their mass models constructed from stellar photometry and high-resolution HI 21cm observations \citep{de_Blok_2001,Banerjee2010}. In contrast, dark matter dominates the disk dynamics of ordinary spirals only at larger radii. See, for example, \cite{Banerjee2008}. Calculating the 2-component disk stability parameter $\mathrm{Q_{RW}}$ proposed by \cite{Romeo_2011} of a sub-sample of LSBs from \cite{de_Blok_2001}, \cite{garg2017origin} showed that the low star formation rates in LSBs could be explained on the basis of the high dynamical stability of their disks against local, axi-symmetric perturbations. Using the 2-fluid disk stability parameter of \cite{Jog1996}, \cite{ghosh2014} found that the galactic disk of the LSB superthin galaxy UGC7321 is stable against local, axisymmetric perturbations as well as local, non-axisymmetric perturbations. \cite{Ghosh2015} studied the response of an LSB galactic disk to global, non-axisymmetric perturbations, and found that the dark matter halo does not play any significant role in suppressing the global instabilities. In fact, they attributed the lack of strong spiral features in LSBs to their sparse environment. However, according to Galaxy zoo2, the bar fraction in LSBs is $\sim$ 0.2 whereas the same in HSBs is $\sim$ 0.3 (see \cite{sodi}) . Therefore, the prevalence of non-axisymmetric features is not that insignificant in LSBs as compared to HSBs. Using N-body simulations, \cite{mihos1996} showed that LSBs are quite stable against the formation of bars or non-axisymmetric instabilities. They also used the analytical parameter introduced by \cite{Goldreich1978} to show the stability of LSBs against a global non-axisymmetric mode. Using N-body simulations, \cite{chiueh} argued that a dynamical coupling between the disc and the dark matter halo of high angular momentum low surface brightness galaxies may drive disk density waves via the co-rotation resonance even in disks with a high value of the Toomre Q parameter. Further, the halo scaleheight was found to regulate the pitch-angle of the spiral arm, with more loosely-would arms driven by smaller scaleheights. Further, hydrodynamical simulations done by \cite{Mayer2004}  indicated that the LSBs are stable against bar formation, and attributed the same to the lack of their self-gravity. Using TNG100, \cite{perez-Montano2022} shows that the high values of specific angular momentum of the stellar disk and the spin of the dark matter halo in LSBs are responsible for their extended nature, and also leading to the absence of massive central black holes. Thus, analytical and numerical studies have shown that LSBs are stable against both local axisymmetric and non-axisymmetric instabilities. Yet, we observe these galaxies are not devoid of spiral features. The origin of spiral structure in LSBs is therefore a puzzle. These galaxies are often found in void or isolated environments, and so the spirals must be formed from self-excited mechanisms. \\

\noindent In several numerical simulation studies, recurrent spiral activity is seen in unbarred isolated disk galaxy. This spiral activity fades over time due to the random stellar motion in the disk, which heats up the stellar disk further \citep{SellwoodCarlberg1984}. In fact, N-body simulations always form a transient spiral, which does not obey density wave theory. Interestingly, \cite{Sellwood2011} found that long-lived spiral modes are not reproducible in simulations. \cite{SellwoodCarlberg2014} simulated recurrent transient spiral patterns, and showed that a recurrent global spiral structure can be explained by the superposition of several modes in isolated, unbarred disk galaxy models.\\

\noindent In this paper we try to understand the origin and evolution of the observed spiral features in two typical LSBs, F568-VI and F568-01, using theoretical models of swing amplification and density waves, as well as N-body + hydrodynamical simulation. The organization of the paper is as follows: In \S 2, we describe the models we use for the galactic spiral arms,  in \S 3 the target galaxies, in \S 4, the input parameters for the different models followed by the results and the conclusions in \S 5 and 6 respectively.\\

\begin{table}
	\centering
	\caption{Physical Properties of F568-V1 \& F568-01:}
	\label{ObjDescpn}
	\begin{tabular}{lcc} 
           \hline
           RA & $\rm{161.259^{\circ}}$ & $\rm{156.526^{\circ}}$  \\
           DEC& $22.054^{\circ}$ &$ 22.434^{\circ}$   \\
           Distance(Mpc) &     90   & 101\\
           $\mathrm{Vrot(kms^{-1})}$ \footnote{Asymptotic rotational velocity}& 99.6 & 100.9 \\
           $\mathrm{M_{HI}(M_{\odot})}$ \footnote{Mass of HI disk}& 6.32E+09 & 6.45E+09 \\
           $\mathrm{L_B(L_{\odot})}$ \footnote{Luminosity of stellar disk in B-band}& 2.77E+09 & 3.12+09 \\
           Inclination & $46.4^{\circ}$ & $31.9^{\circ}$\\
           Position angle & $48.1^{\circ}$ &$57.0^{\circ}$\\
		\hline
	\end{tabular}
        \label{PhysP}
\end{table}

\section{Modelling of Galactic Spiral Arms}

\subsection{Local Spiral Arms}

\begin{table}[]
    \centering
    \caption{Input parameters: Local Spiral Arms \& Global Mode study}
    \begin{tabular}{lcc}
    \hline
    Input Parameters & \\
    \hline
    Gas disk \footnote{parameters in $\mathrm{\Sigma_{HI} = A e^{-\frac{(r-m_1)^2}{s_1^2}}+B e^{-\frac{(r-m_2)^2}{s_2^2}}}$}& F568-VI & F568-01 \\
    \hline
        A & 1.7 $\rm{M_\odot pc^{-2}}$ & 3.97  $\rm{M_\odot pc^{-2}}$ \\
        $m_1$& 7.9 kpc & 4.45 kpc\\
        $s_1$ & 2.3 kpc & 3.98 kpc\\
        B & 3.5 $\rm{M_\odot pc^{-2}}$ & 3.98 $\rm{M_\odot pc^{-2}}$\\
        $m_2$ &2.9 kpc & 4.46 kpc\\
        $s_2$  & 4.3 kpc & 3.98 kpc\\
	\hline
        Stellar disk & \\
        \hline	
        $\Sigma_0$ \footnote{central stellar surface density in B-band}  & $\mathrm{34.20 \ M_\odot \ pc^{-2} }$ & $\mathrm{21.58 \ M_\odot \ pc^{-2} }$ \\
        $R_d$ \footnote{Exponential stellar disk scale length}& $\mathrm{3.1 \ kpc}$ & $\mathrm{5.3 \ kpc} $\\
	\hline
        Dark matter halo& \\ 
        \hline
        $\rho_0$& 0.141 $\mathrm{M_\odot  pc^{-3}}$ & 0.097 $\mathrm{M_\odot  pc^{-3}}$ \\
        $R_c$  & 1.41 kpc & 2.11 kpc  \\
 \hline
        a & $\mathrm{115 \ km s^{-1}}$&$\mathrm{137 \ km s^{-1}}$ \\
	b \footnote{parameters from fit to the rotation curve ($\rm{a \ (1-exp(-r/b))}$)}& $\mathrm{2.19 \ kpc}$& $\mathrm{2.8 \ kpc}$ \\
	\hline
        $Q_s$ \footnote{Toomre Q for stars at $2 
 \ R_d$}& 10.1 & 5.5\\
        $h_z$ \footnote{stellar scale height used in deriving vertical velocity dispersion}& $\rm{R_d/6}$ & $\rm{R_d/6}$\\
        $\sigma_z(0)$  \footnote{central stellar velocity dispersion} & $32 \ \mathrm{kms^{-1}}$& $20.19\ \mathrm{kms^{-1}}$ \\
        $\alpha$ \footnote{scale length in stellar velocity dispersion profile}& 2.0 & 2.0\\
        $Q_g$ \footnote{Toomre Q of gas disk at $2 
 \ R_d$ }& 11.2 & 6.6 \\
        $\sigma_{HI}$ \footnote{velocity dispersion of gas disk}& 7 \ $\mathrm{kms^{-1}}$ & 7 \ $\mathrm{kms^{-1}}$\\
		$c_{z}(0)$ \footnote{central stellar vertical velocity dispersion for global mode study}   & $\mathrm{32 \ km \ s^{-1}}$ &  $\mathrm{20.19\ km \ s^{-1}}$\\
            $\rm{R_{OUT}}$ \footnote{truncation radius of stellar surface density and velocity for global mode study} & 15.6 kpc& 26.5 kpc \\
            \hline 
    \end{tabular}
    \label{InpSAmp}
\end{table}

\begin{table}
	\centering
	\caption{Input parameters for the RAMSES simulation}
	\label{InpRAMSES}
	\begin{tabular}{lccc} 
		\hline
             Input Parameters & F568-VI & F568-01 \\
            \hline
		Virial velocity, V200 & 49 $\rm{kms^{-1}}$&112.1$\rm{kms^{-1}}$  \\
		Spin parameter, $\mathrm{\lambda}$ & 0.02 &  0.02 \\
		DM halo mass fraction& 0.95 & 0.95\\
		DM concentration & 7.55 & 13.4 \\
		Scale length & 8.55  kpc& 8.55  kpc\\
		Axis ratio \ (b/a)  & 0.7& 0.7\\
		No of Dark matter particles& 2E+5&2E+5\\
		\hline
	    Stellar disk mass fraction& 0.02 & 0.02\\
            Stellar disk scale length, $R_d$& 3.1 kpc& 5.3 kpc \\
            Stellar disk scale height& $\mathrm{R_d / 10}$& $\mathrm{R_d / 10}$ \\
		No of star particles& 5E+5&  5E+5\\
            Stellar disk Toomre Q& 1.5& 1.5 \\
		\hline
		Gas disk mass fraction& 0.02 & 0.02\\
            Gas disk scale length, $\mathrm{R_d}$& 50.0  $kpc$& 7  $kpc$\\
            Gas scale height& $\mathrm{R_d / 20}$ & $\mathrm{R_d / 20}$\\
		Gas disk Toomre Q& 1.0 & 1.0\\
            No of gas particles& 1E+5& 1E+5\\
		\hline
		Star formation efficiency, $\mathrm{\epsilon_*}$ & 0.05& 0.05 \\
		Supernova mass fraction, $\mathrm{\eta_{SN}}$ &0.05 & 0.05\\
		\hline
	\end{tabular}
\end{table}

\noindent Galactic disk could be stable against local, axi-symmteric perturbations as indicated by a high value of the Toomre Q parameter \citep{toomre1964gravitational} and yet unstable against the growth of local, non- axisymmetric instabilities. The growth of these local, non-axisymmetric instabilities was first studied by \cite{Goldreich1965}, and \cite{Julian1966} proposed the swing amplification mechanism to explain the origin of spiral arms in stellar disk of galaxies in response to them. The term swing amplification was coined by \cite{Toomre1981} as the wave gets amplified as it swings from a leading to a trailing position due to differential rotation of the galaxy. The swing amplification mechanism can only be used to understand the local or flocculent spiral arms and not directly to understand the global spiral structure \citep{Michikoshi2016SA}. Further, N-body simulations with spherical halos produce recurrent and transient spirals which are mostly excited by the swing amplification mechanism \citep{SellwoodCarlberg1984, Baba_2009, Fujii_2011}. \\

\noindent Following \cite{Jog1992}, we model the galactic disk as a 2-fluid disk of stars and gas with zero vertical thickness, differentially-rotating, and gravitationally-coupled to each other, and also in the force-field of a \emph{spherical}, pseudo-isothermal  dark matter halo, and study the linear response of the disk to local, non-axisymmetric perturbations. We present here the final coupled differential equation describing the growth of $\mathrm{\theta_i=\frac{\delta \mu_i}{\mu_{0i}}}$, a dimensionless quantity which represents the ratio of the perturbed to the unperturbed surface density of the $i^{\rm{th}}$ component ($i$ = stars, gas) as a function of $\tau$, 
a dimensionless measure of time in a sheared co-ordinate system. Here the sheared co-ordinate system is defined as $x'=x, y'=y-2Axt, z'=z, t'=t$ where $A = \frac{1}{2}(\frac{V_{\rm{rot}}}{R} - \frac{d{V}_{\rm{rot}}}{dR})$.
$\tau$ is given by $\tau=2At'- k_x/k_y$ where $k_x$ and $k_y$ are respectively the wave numbers in the $x$ and $y$ directions and is hence a measure of the dynamical time scale of the disk. \\ 

\begin{equation}
\begin{split}
\left(\frac{d^2\theta_i}{d\tau^2} \right)-\left(\frac{d\theta_{i}}{d\tau}\right) \left(\frac{2\tau}{1+\tau^2}\right) \\
+\theta_s \left[ \xi^2 + \frac{2(\eta-2)} {\eta(1+ \tau^2)}+\frac{(1+\tau^2)Q_i^2(1-\epsilon)^2\xi^2}{4\chi^2}
\right]
\\
=\frac{\xi^2}{\chi}(1+ \tau^2)^{1/2}\left[\theta_s(1-\epsilon)+\theta_g\epsilon\right]
\end{split}
\end{equation}

\noindent Here $Q_i$ denotes the Toomre Q parameter for the $i^{th}$ disk component, where i = stars and gas, with ${Q_i}>1$ indicating that the $i^{\rm{th}}$ galactic disk component is stable against local, axi-symmetric perturbations. $\eta$ is the logarithmic shearing rate in the galactic disk $\rm{(\eta = (R/\Omega)/(d\Omega/dR))}$
and $\chi$ the wavelength of non-axisymmetric perturbation $\lambda$ in units of the critical wavelength $\rm{{\lambda}_{\rm{crit}}}$; $\eta<1$ and $\eta>1$ corresponds to the rising  and the falling part of the rotation curve respectively while $\eta=1$ represents a flat rotation curve ($\xi^2=2$). Here $\xi^2 = (4-2\eta)/\eta^2$. Further, $\epsilon$ is the ratio of surface density of gas disk to that of the stellar disk given as $\Sigma_g / \Sigma_s$  and $\chi$  is the wavelength of axi-symmetric perturbation  in units of  $\rm{\lambda_{crit}}$. The above system of equations may be solved if the above 2-component disk is stable against local, axi-symmetric perturbations, which is given by the following criterion \citep{JogSolomon1984}:

\begin{equation}
\begin{split}
\frac{(1 - \epsilon)}{ \chi (1+(Q_{s}^{2} (1-\epsilon)^{2})/4 \chi^{2}))} \\
+\frac { \epsilon }{ \chi(1+(Q_g^2 \epsilon^2 /4 \chi^{2}))}<1 
\end{split}
\end{equation}

\noindent \textbf{Solution of the equations:} The coupled, second-order, ordinary differential equations given by Equation (1) are solved iteratively using fourth order Runge-Kutta method by imposing suitable initial conditions. There are four possible sets of initial conditions for ($\theta_s$,$\dot{\theta_s}$,$\theta_g$,$\dot{\theta_g}$) : $(1,0,0,0),(0,1,0,0),(0,0,1,0),(0,0,0,1) $. The magnitude of amplification also depends on the choice of the initial value of $\tau$. Therefore, $\tau$ is also varied with each initial condition and the combination which leads to the maximum amplification is chosen. The initial value of $\tau$ was varied and the value which gives maximum amplification is retained. If not, the $\tau$ was chosen for a greater maximum amplification factor  MAF (=$(\theta_i)_{\rm{max}}/(\theta_i)_{\rm{ini}}$) of star or gas. The values of the input parameters like $\eta$ and $\xi$ may be directly calculated from the observed rotation curve. $\chi$ is varied in the range 1 - 3 as the swing amplification mechanism becomes ineffective outside this range of $\chi$ \citep{Toomre1981}; the $\chi$ value which results in the maximum amplification is determined by trial and error method. 

\subsection{Global Spiral Modes}

\noindent Observational studies suggest that the galactic spiral arms are stable against the differential rotation of the galaxy and rotate as a rigid body with a constant pattern speed. This indicates that the spiral arms cannot be material arms as otherwise they would have wound up resulting in tightly-wound spirals only, just in a few dynamical times. \cite{Linblad1948,Linblad1959} and \cite{ Lin1964} found a way out of the winding dilemma by modeling the spiral structures as stationary, density waves rather than winding material spiral arms. Following \cite{Korchagin2000}, we model the galaxy as a 1-component fluid disk obeying the Euler equation, and responding to the gravitational potential of its self-gravity as well as the external force-fields of the gas disk and the dark matter halo, as governed by the Poisson equation. Perturbing the Euler equation by a global, non axi-symmetric instability of the form
\begin{equation}
f(r)+f_1(r)e^{im\phi-i\omega t}
\end{equation}

with $f(r)$ denoting any unperturbed quantity, and $f_1(r)$ the corresponding amplitude of the perturbed quantity, $m$ the azimuthal wave number and $\omega$ is the complex frequency of the perturbation. The linearized equation is then given by

\begin{equation}
\begin{split}
\frac{d^2}{dr^2}(w_1+\psi_1)+A \frac{d}{dr} (w_1+\psi_1)+ \\
B(w_1+\psi_1) - \frac{D}{c_r^2} (w_1) = 0 \\
A = \frac{2m+1}{r} 
+\frac{1}{\Sigma} \frac{d \Sigma}{dr} -\frac{1}{D} \frac{dD}{dr} \\
B = \frac{m}{r} \bigg[ \left( \frac{1}{\Sigma}\frac{d \Sigma}{dr} -\frac{1}{D} \frac{dD}{dr} \right)    
\bigg(1 - \frac{2 \Omega}{ \omega - m \Omega} \bigg) \\
- \bigg( \frac{2}{ \omega - m \Omega} \frac{d \Omega}{dr} \bigg) \bigg] \\
D = \kappa^2 - (\omega - m \Omega)^2 
\end{split}
\end{equation}

\noindent where $\Omega$ is the angular velocity of the disk, $\Sigma$ the stellar surface density and $\kappa$ the epicyclic frequency derived from rotational velocity profile as $\mathrm{\frac{2V}{r^2}\frac{d}{dr}\big(\frac{V}{R}\big)}$.
$\omega_1$ is the perturbed enthalpy, $\Sigma_1$ is the perturbed surface density, and the perturbed gravitational potential  is given as

\begin{equation}
\begin{split}
\tilde{\psi_1}(r) = -2 \pi \frac{\Gamma(m + 1/2)}{\Gamma(1/2) \Gamma(m+1)}  
\\
\bigg[\int_0^r \Sigma_1(r')\frac{r'}{r}^{2 m+1}  
F \left(0.5, m+0.5, m+1, \frac{r'}{r}^2 \right)
dr'+ \\
\int_r^{R_{out}} \Sigma_1(r') F\left(0.5, m+0.5, m+1, \frac{r'}{r}^2\right)dr'\bigg] 
\\
\end{split}
\end{equation}

\noindent Here $F$ denotes the hypergeometric function. The surface density and the radial velocity dispersion are respectively modelled as
\noindent

\begin{subequations}
\begin{equation}
\Sigma    = \Sigma_0\rm{exp}(-r/h_\sigma)((1- r/R_{OUT})^2)^5 
\end{equation}

\begin{equation}
c_z  = c_{z}(0)\rm{exp}(-r/(2h_\sigma))((1- r/R_{OUT})^2)^{2.5}
\end{equation}
\end{subequations}

\noindent where $\Sigma_0$ is the central surface density, $h_\sigma$ the radial disk scale length, $\rm{c_{z}(0)}$ the central vertical dispersion, and $\rm{R_{OUT}}$ the truncation radius. The boundary conditions are obtained by taking the radial stellar velocity dispersion $\rm{c_r = 0}$ at the boundaries. The equations are thus reduced to a matrix equation and the eigen frequency spectrum is found, following \cite{Adams1989}. The growth time and the pattern speed of the mode is given by as $2 \pi/Im(\omega)$ and by $2 \pi/Re(\omega)/m$ respectively.

\subsection{N-body + Hydrodynamical Simulations}
\noindent The galactic disk is modelled as a 2-component system of stars and gas in an NFW halo \citep{Navarro1996}, all the three components being gravitationally-coupled to each other. Both the stellar and the gas surface density radial distributions were taken to follow exponential profiles. The radial and vertical velocity dispersion are regulated using Toomre Q and vertical scale height respectively. The physical properties like asymptotic rotational velocity and surface density of the disk components are used as constraints to generate the initial conditions. We use a publicly-available code DICE \citep{VPerret2016} to generate initial conditions for the galaxy in equilibrium. The initial conditions are generated using Lagrangian particles whose distributions are built using a Metropolis-Hasting MonteCarlo Markov Chain algorithm \citep{Metropolis1953}. Simulation is run with $\mathrm{2\times10^5}$ dark matter halo particles, $5\times10^5$ stellar disk particles and $\mathrm{1\times10^5}$ gas particles following \citet{Mayer2004, Sellwood2019}. We evolve the galaxy model using the publicly available code RAMSES \citep{Teyssier2002}. The code use Adaptive Mesh Refinement (AMR) technique and tree-based data structure which allows recursive grid refinements. The hydrodynamical solver is based on second-order Godunov method, which computes the thermal history of the fluid component with high accuracy. All the plots were generated using publicly-available software pynbody \citep{pynbody1}.

\section{Targets: FGC568-VI \& FGC568-01}

\noindent We choose two prototypical LSBs, F568-V1 and F568-01, studied by \cite{de_Blok_2001} for our study. Both are seen face-on with angles of inclination $\sim 46.4^\circ$ and $\sim 31.9^\circ$ respectively, with the corresponding distances being 90 Mpc and 101 Mpc. Astrometric parameters of the galaxy are obtained from Hyperleda\footnote{http://leda.univ-lyon1.fr/} \citep{makarov14} and are quoted in \autoref{ObjDescpn}. The respective asymptotic rotational velocities $V_{\rm{rot}}$ are 99.6 and 100.9 kms$^{-1}$, confirming that their total dynamical mass are intermediate between that of dwarfs and ordinary galaxies like the Milky Way. The masses of the HI disk and the stellar disk are both of the order of $10^9$ $M_{\odot}$, indicating that the self-gravity of the stars and the gas are equally important in regulating the disk dynamics. The physical properties of the galaxies are presented in \autoref{PhysP}.

\section {Input Parameters}
\noindent The radial profile of both the stellar surface density in $B$-band, as well as the gas surface density, were taken from \cite{deBlok_1995}.
The rotation curve was taken from \cite{de_Blok_2001}.

\subsection{Local Spiral Arms \& Global Spiral Mode}
\noindent For the Swing Amplification study, the observed gas surface density profile was fitted with a linear superposition of two Gaussian profiles, but not centered at zero: $\mathrm{\Sigma_{HI} = A e^{-\frac{(r-m_1)^2}{s_1^2}}+B e^{-\frac{(r-m_2)^2}{s_2^2}}}$. See, for example, \cite{Patra2013}. The surface density of the stars was fitted with an exponential profile $\mathrm{\Sigma_0 exp(-R/R_d)}$. For the Global Mode study, the gas surface density profile was not required. For the stellar surface density, $R_{OUT}$ is taken to be 5 $R_d$. $R_{OUT}$ indicates the radius beyond which the stellar surface density becomes negligible or zero. Assuming that the stellar surface density “zero” at $R_{OUT}$ is necessary to comply with the set of suitable boundary conditions of the problem as formulated \citep{Korchagin2000}. Our conclusions remain unchanged for any value of  $R_{OUT}$ equal to or larger than 5 $R_d$.  In both cases, the rotation curve was fitted with $\mathrm{a(1-\rm{exp}(-r/b)}$ where $a$ and $b$ are constants.

\noindent The stellar velocity dispersion for our sample galaxy was not available from spectroscopic observations, and was therefore analytically modelled. We use the 2-component model of gravitationally-coupled, stellar and gas disks, in the force-field of the dark matter halo and in vertical hydrostatic equilibrium as constrained by the observed stellar and gas scale height to model this \citep{Narayan2005}. The dark matter parameters were taken from the mass models with pseudo-isothermal (PIS) halo \citep{de_Blok_2001}. Since our LSB is not edge-on, its stellar and the HI vertical scale height are not directly measured. We assume that the mean scale height of stellar disk to be  $\sim$ $R_{d}/6$  following the scaling relation determined from the study of a sample of edge-on spirals which says that the stellar scale height lies between $R_{d}/5-R_{d}/7$ \citep{VanderSearle1981}. The HI scale height was taken to linearly vary between $\sim$ 0.2 to 1 kpc between $R$ = 0 to 0.6 $R_d$ following the trend observed in a sample of edge-ons by \cite{Obrien2010}(See Figure 25 of their paper). We further assume the radial variation of the vertical, stellar dispersion profile as  ${\sigma}_{z}(R)$= ${\sigma}_{z}(0)$ $\rm{exp}(-R/{{\alpha}R_d})$, where $\sigma_{z}(0)$ is the central vertical stellar dispersion value and ${\alpha} R_d$ the scale length for the fall-off of the dispersion, $R_d$ being the exponential disk scale length of the stellar disk. This assumed radial profile of ${\sigma}_{z}(R) $ is due to \cite{VanderSearle1981}, who found this to be well-consistent with the flat, radial profiles of the stellar scale heights as observed in a sample of edge-on galaxies. Interestingly, we note there is hardly any notable variation of ${\sigma}_z$  with $h_z$. This is due to the fact that dispersion ${\sigma}_z$ varies as $\sqrt{(h_z)}$ , at least in the one component model, and so the dependence of ${\sigma}_z$ on the assumed value of $h_z$ is weak. Besides, ${\sigma}_{HI}$ was assumed to remain constant with $z$, which is a reasonable assumption given a thin vertical structure of the stellar disk.  For HI dispersion ${\sigma}_{HI}$, we consider it to be constant at all $R$ at a canonical value of 7 $\rm{kms}^{-1}$. \\

\noindent We first obtain the vertical stellar dispersion and then we multiply it by a factor 2 obtain the central radial stellar dispersion,  0.5 being the ratio of the vertical-to-radial stellar dispersion observed at the solar neighbourhood (See, for example, \cite{Binney2008}). However,  recent studies have shown that this ratio may take a range of values with a value of $\sim$ 0.3 more appropriate for late-type galaxies like the LSBs \citep{Gerssen2012}. Therefore, our choice of a vertical-to-planar stellar velocity dispersion ratio of 0.5 is rather conservative. In case of the Swing Amplification study, the radial velocity dispersion values were required to calculate the Toomre Q values for the stellar and the gas disks as $Q_i = \kappa \sigma_R/(\pi G \Sigma)$, symbols having usual significance. In case of the Global Mode study, only the radial stellar velocity dispersion was required, and was used directly as an input parameter. All the input parameters discussed above are listed in \autoref{InpSAmp}. \\

\subsection{N-body + Hydrodynamical simulations}

\noindent DICE does not allow superposition of two Gaussian profiles for the gas disk. So, an exponential profile with a large enough radial scale length, mimicking the average gas surface density value and giving the same total mass was fitted to generate in the initial conditions.  We used a Navarro-Frenk-White (NFW) dark matter halo for our simulations. This was preferred over a pseudo-isothermal halo to comply with the observed shape of the rotation curve. The parameters were taken from the NFW mass models of \cite{de_Blok_2001}. In the ILLUSTRIS TNG simulation, the median value of intermediate-to-major axes ratio $\rm{b/a}$ for 14,000 dark matter halos in mass range $10^{11} - 10^{14} m_\odot$ is 0.8 \citep{Chua2019}. Also ILLUSTRIS-dark simulation  without baryons results in halos with 0.7 as median $b/a$. We initialize dark matter halo in our simulation with a similar value. The dark matter halo spin is given by $\mathrm{\lambda = j /\sqrt{2} V_{vir} R_{vir}}$ where $\rm{V_{vir}}$ and $\rm{R_{vir}}$ are the velocity at the virial radius and the virial radius respectively \citep{Bullock2001}. Also dark matter halos with higher spin promotes bar formation in the stellar disk \citep{Saha2013, Weinberg1985}. \cite{Tonini2006} showed $\lambda$ peaks at 0.025 for dark matter halos hosting spiral galaxies. LSBs are found in fast rotating halos with large angular momenta \citep{1997Jiminez, 2003Boissier}. We have chosen spin parameter as 0.02  which is similar to that of spiral galaxies. Q is set in range of  $1-2$ which is ideal for the growth of spirals by swing amplification \citep{Goldreich1965,Julian1966,Toomre1981}. In \autoref{InpRAMSES}, we present the input parameters for the RAMSES simulations.
 
\section{RESULTS}

\subsection{Linear Perturbation Analysis}

\subsubsection{Local Spiral Arms}

\begin{figure*}
\begin{center}
\begin{tabular}{c}
\resizebox{150mm}{40mm}{\includegraphics{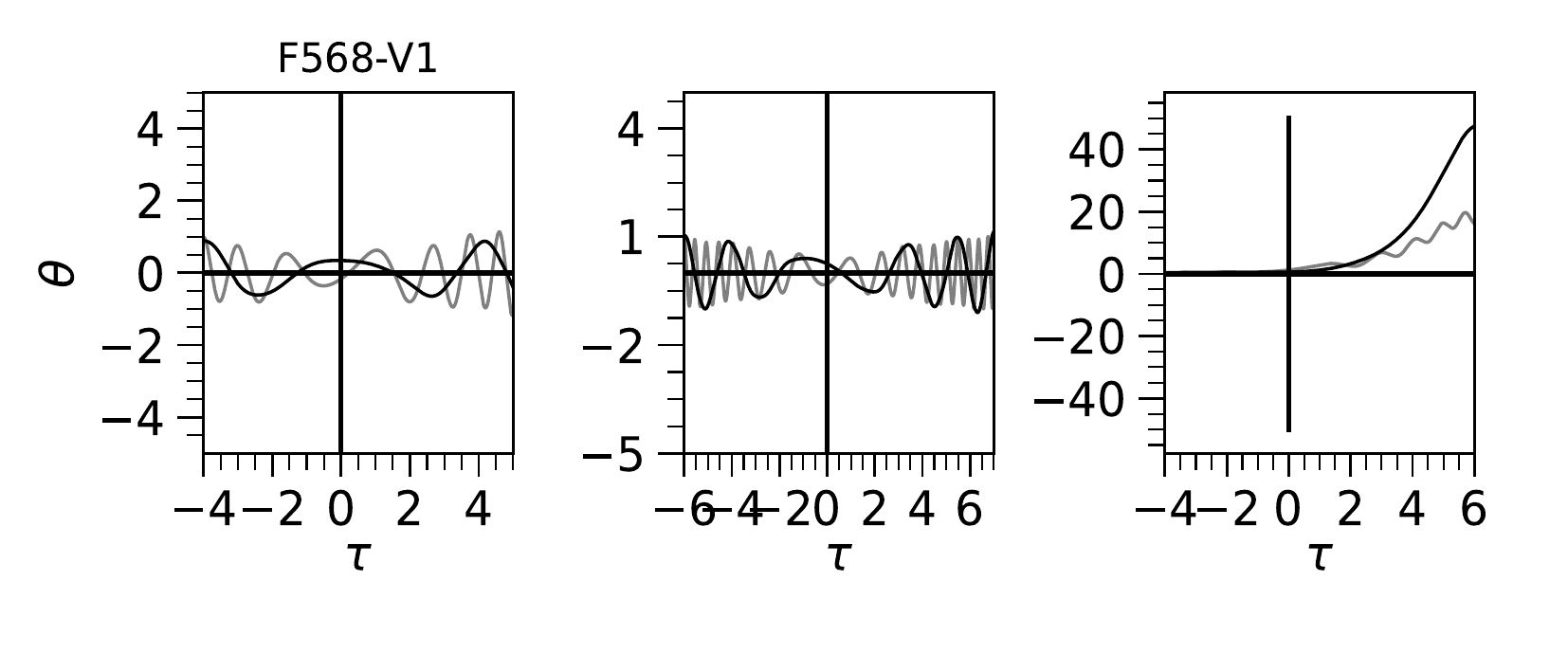}} \\
\resizebox{150mm}{40mm}
{\includegraphics{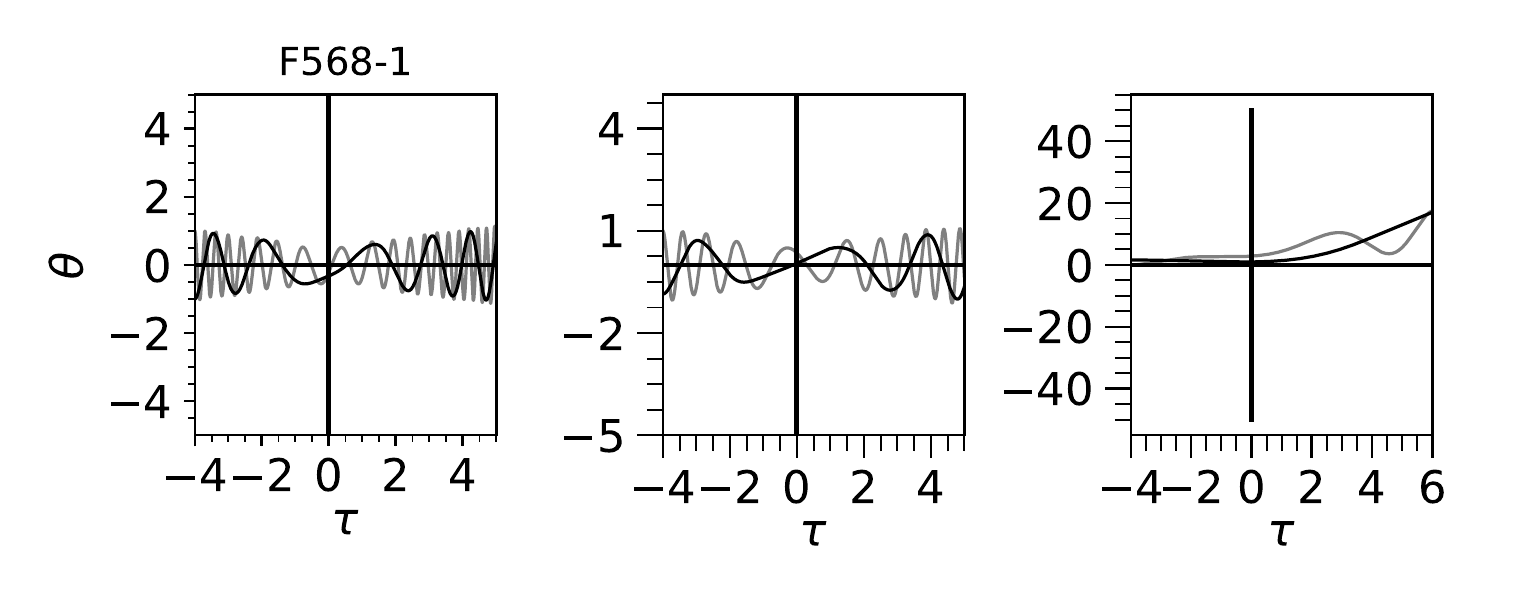}}
\end{tabular}
\end{center}
\caption{Local Spiral Arms: Response of the stellar (grey) and gas (black) disk of a 2-component gravitationally-coupled system of stars and gas when subjected to local, non-axisymmetric perturbations in the total gravitational potential (Left Panel), in the dark matter halo potential only (Middle Panel)  and  the disk potential only (Right Panel).Top panel shows the response of the galaxy F568-V1 and bottom panel depicts the response of F568-1. }
\label{SA}
\end{figure*}

\noindent In \textcolor{red}{\autoref{SA}}, we show that response of the stellar and the gas disk in the gravitational potential of the disk and the dark matter halo (Left Panel), the dark matter halo only (Middle Panel), and the disk only (Right Panel) of our sample galaxies: F568-VI (Top Panel) and F568-01 (Bottom Panel). We find that dark matter plays a crucial role in inhibiting local, non-axisymmetric instabilities that grow by the swing amplification mechanism. \emph{This confirms that the disks of both F568-VI and F568-01 are stable against the growth of local spirals by swing amplification}. The galactic disk is susceptible to non-axisymmetric instabilities if the local disk Toomre Q is in range of 1-2 \citep{Toomre1981}. In this case, both $Q_s$ and $Q_g$ are $\sim$ 10, and one may tend to attribute the absence of swing amplification to the high $Q$ values. However, we checked the same by taking both $Q_s$ and $Q_g$ $\sim$ 1.5, but that too does not lead to the growth of spiral arms. Interestingly, LSBs and other low luminosity galaxies are characterized by patchy and irregular spiral features, which are often understood to be developed by the amplification of the local, non-axisymmetric perturbations by the swing amplification mechanism, among others. \\

\subsubsection{Global Spiral modes}

\begin{figure*}
\begin{center}
\begin{tabular}{cccc}
\resizebox{75mm}{!}{\includegraphics{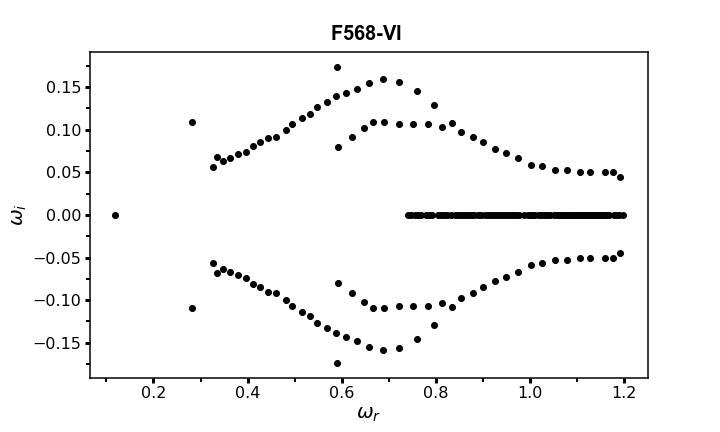}}
\resizebox{75mm}{45mm}{\includegraphics{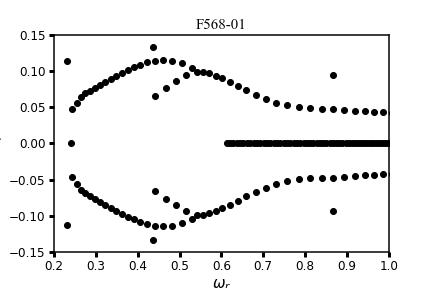}}
\end{tabular}
\end{center}
\caption{Global Mode Analysis: The Eigen Spectrum for LSB F568-V01[left] and F568-01[right]}
\label{GModeRes}
\end{figure*}

\noindent In \autoref{GModeRes}, we present the eigen spectrum of the Global Mode Analysis of the LSB disks of F568-VI and F568-01. The imaginary part determines the growth rate, while the real part gives the pattern speed of the eigenmode. For F568-VI, the real and imaginary values of the most unstable mode are 0.56 and 0.16, respectively. This gives a pattern speed of $\rm{22 \ km s^{-1} kpc^{-1}}$, and a growth-time of $\mathrm{0.78}$ Gyr, which is 4 times the dynamical time of the galaxy ($0.2$ Gyr at $1.5$ $\rm{R_d}$). For F568-01, the real and imaginary values of the same are 0.44 and 0.13, respectively. This indicates a pattern speed of $\rm{20 \ km s^{-1} kpc^{-1}}$ and a growth-time of $\mathrm{0.96}$ Gyr, which is 2.3 times the dynamical time of the galaxy ($0.42$ Gyr at $1.5$ $\rm{R_d}$). The growth times are comparable to the range of growth rates observed by \cite{Korchagin2000}. \emph{This already implies that the LSB stellar disks are susceptible to the growth of global spiral modes, which is also in line with the observations made by Sodi and Garcia (2017)} However, there is a caveat as in this model: the self-gravity of the gas is not included in our study, and gas, being a cold component, may render the disk unstable. Response of a 2-component system of gravitationally-coupled stars and gas in a \textbf{live} dark matter halo to global, non-axisymmetric perturbations is not tractable by analytical models, and one has to resort to N-body + hydrodynamical simulations to study the problem.  

\subsection{N-body + Hydrodynamical simulations using RAMSES: : A Non-Spherical Dark Matter Halo}

\noindent As noted earlier, the initial conditions for our simulations are generated using DICE. We emphasize here that the initial conditions thus generated by DICE are indeed in equilibrium. This is because these are constrained by the observed stellar and gas surface density profiles, and rotation curve. In addition, we consider the different components of the rotation curve as was obtained from the mass model, as constraints.  We further check that the star-gas-halo system was still in equilibrium when the simulation start by confirming that the same constraints are complied with for the first several epochs. \\ 

\begin{figure*}
\begin{center}
\begin{tabular}{cc}
\resizebox{75mm}{!}{\includegraphics{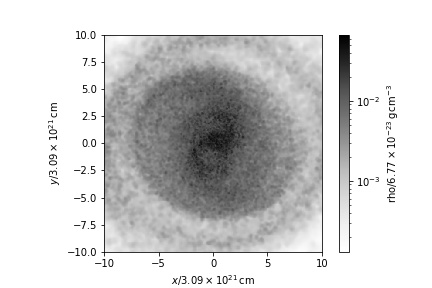}} 
\resizebox{75mm}{!}{\includegraphics{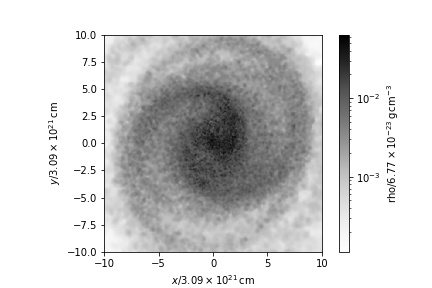}} \\
\resizebox{75mm}{!}{\includegraphics{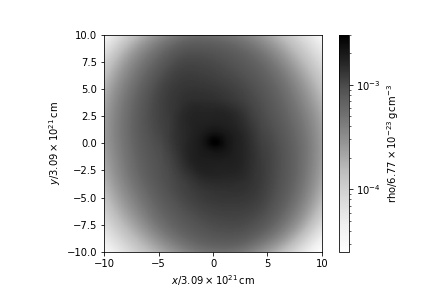}} 
\resizebox{75mm}{!}{\includegraphics{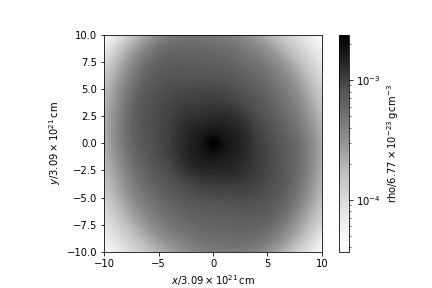}} \\
\end{tabular}
\end{center}
\caption{N-body + Hydrodynamical Simulations of FGC568-VI: Snapshots from volume density at $0.78$ Gyr (Left) and $1.34$ Gyr (Right) of the stellar disk (Top) and gas disk (Bottom). Spiral features begin to appear in the stellar disk by 0.78 Gyr. The pitch angle of the simulated spiral matches with the observed SDSS image at 1.34 Gyr.}
\label{SimSnapshot1}
\end{figure*}

\begin{figure*}
\begin{center}
\begin{tabular}{cc}
\resizebox{75mm}{!}{\includegraphics{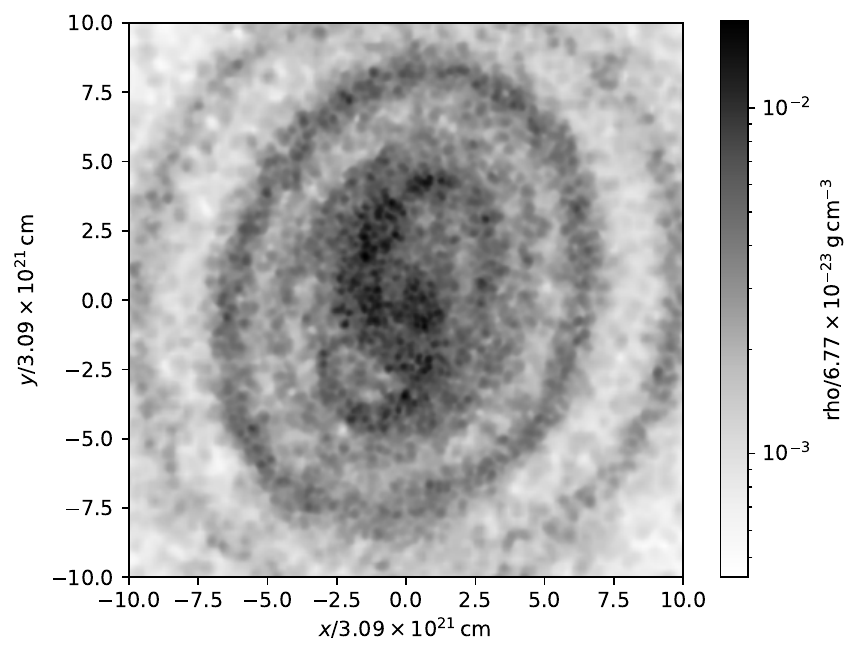}} 
\resizebox{75mm}{!}{\includegraphics{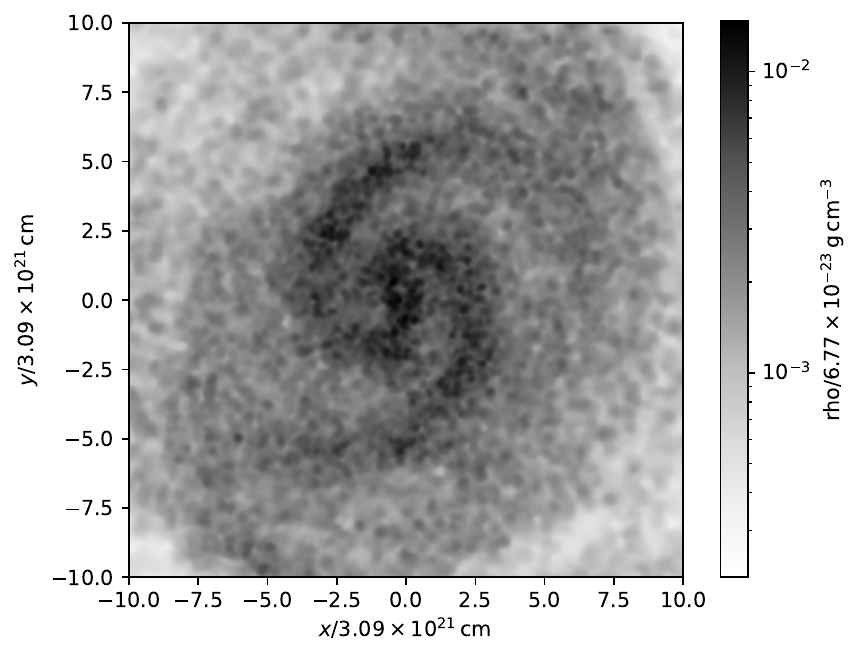}} \\
\resizebox{75mm}{!}{\includegraphics{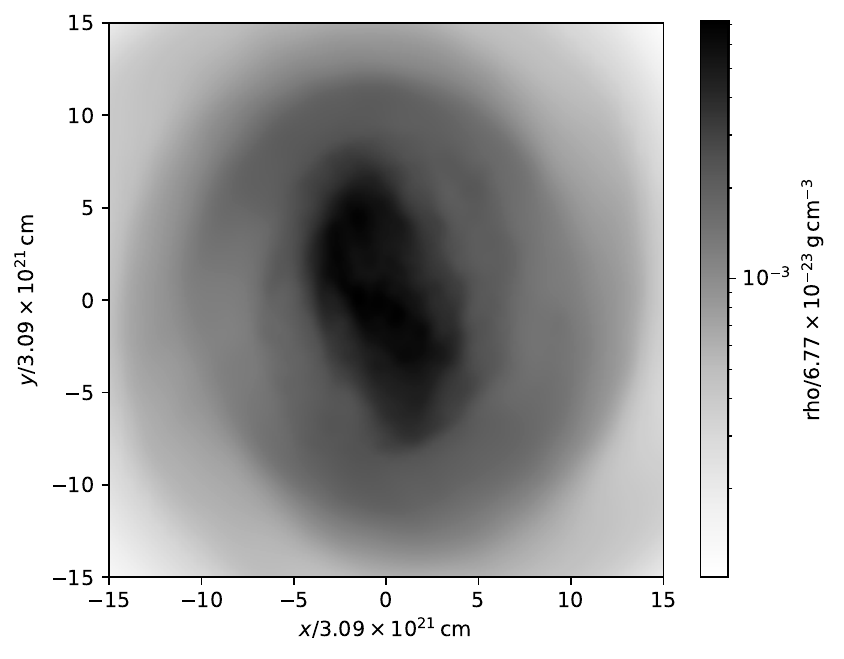}} 
\resizebox{75mm}{!}{\includegraphics{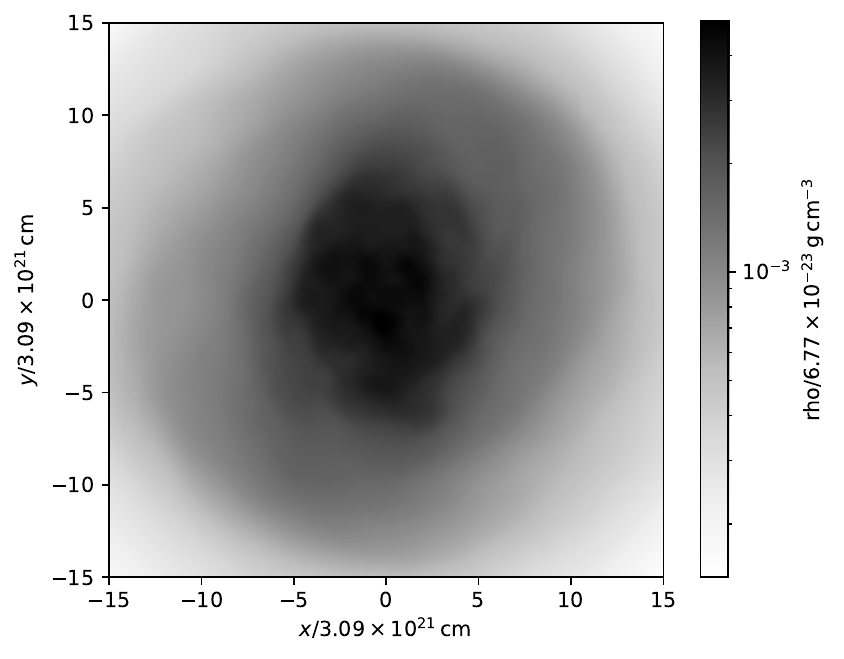}} \\
\end{tabular}
\end{center}
\caption{N-body + Hydrodynamical Simulations of FGC568-01: Snapshots from volume density at $0.5$ Gyr (Left) and $1.4$ Gyr(Right) of the stellar disk (Top) and gas disk (Bottom). Spiral features begin to appear in the stellar disk by 0.46 Gyr. The pitch angle of the simulated spiral matches with the observed SDSS image at 1.4 Gyr.}
\label{SimSnapshot2}
\end{figure*}

\noindent \autoref{SimSnapshot1} shows the snapshots of the stellar \textbf{(top panel)} and the gaseous disks \textbf{(bottom panel)} of FGC568-VI as determined from the volume density at $0.78$ Gyr and $1.34$ Gyr respectively. At $0.78$ Gyr, the spiral wave just begins to appear; at $1.34$ Gyr, the pitch angles of the simulated and the observed LSB match the best. The spiral arms extend over $\sim$ 10 kpc, which is about 3 disk scale lengths. Similarly, in \autoref{SimSnapshot2}, we present the snapshots of the stellar and the gaseous disks of FGC568-01 as obtained from the volume density at $00.46$ Gyr and $1.4$ Gyr, respectively.  The spiral arms extend over $\sim$ 10 kpc, which is about 2 disk stellar scale lengths. For FGC568-VI, the oblate halo triggers spiral perturbations, which begin to develop around $0.78$ Gyr, which is $\sim 4$ times the dynamical time of the galaxy. Incidentally, this is also the growth time of the most unstable global, non-axisymmetric mode, as discussed in \S 5.1.2. We note that the spiral activity in the stellar disk persists for approximately 28 dynamical times ($\sim 2.1 Gyr$). For FGC568-01, spiral features begin to appear around $0.5$ Gyr, which is smaller than the growth time of the most unstable mode in this galaxy (\S 5.1.2). This possibly underscores the limitation of the analytical model of global mode analysis in studying spiral structures. Here, the spiral arm survives for more than $3$ Gyrs, which is about $7$ dynamical times. \emph{Interestingly, in either of the galaxies, the gas disks do not develop spiral features, which is a puzzle. See \S 5.2, last paragraph, for a discussion.}\\

\noindent \emph{The oblate dark matter halo:} The dark matter halo is often modelled as spherical in shape with either the pseudo-isothermal (PIS) or Navarro-Frenk-White (NFW) density profile. In the first attempt, we initialized our simulations with an NFW dark matter halo, spherical in shape and evolved the system for 10 Gyrs. However, no spiral features were observed to develop in the LSB disk. Therefore, an oblate halo with vertical-to-planar axes ratio $\rm{c/a = 0.7}$ was used later. Milky Way type galaxies in cosmological zoom-in simulations a shows a triaxial halo with median $c/a$ = 0.9  (\citep{Preda2019}. However, an oblate halo with $c/a$ = 0.9 failed to produce a long-lived, global spiral of reasonable strength in our stellar disks. \emph{The quadrupolar potential of the dark matter halo plays the key role in triggering and sustaining spiral features in the LSB disks of both F568-VI and F568-01.} Our results are in compliance with \cite{Masset2003}, who showed that a triaxial halo could induce spiral in gas disk of a compact dwarf galaxy NGC 2915, using hydrodynamic simulations. \\ \\

\noindent \emph{Choice of Q value for the stellar disc:} We first use $Q_s = 10$ for FGC568-VI and  $Q_s = 5.5$ FGC568-1, as constrained by an assumed stellar scaleheight of $Rd/5$ (See \S 4.1). However, no spiral features were formed. Next, we successively tried $Q_s$ values corresponding to smaller values of the assumed scale height. We note $R_d/20$ is the scale height of one of the thinnest galaxies observed \citep{Aditya2022}, and the corresponding $Q_s$ is 3.9. However, we find that spiral features are not triggered even in a disk with $Q_s$ as low as 1.5. until a sufficiently oblate potential of the dark matter halo, with $c/a = 0.7$ is used. \emph{Therefore, we conclude that the oblate shape of the dark matter halo plays the most crucial role in driving spiral features in the LSB disks.} We may note here that \cite{chiueh} argued that the disc-halo interaction in LSBs may trigger disk density waves even in disks with a high value of the Toomre Q parameter. However, their simulations were N-body-only simulations, and hydrodynamical effects were not taken into account. In summary, we did several simulations for all possible combinations of the following c/a and stellar Q values: $c/a$ = 1, 0.9, 0.8, 0.7, stellar $Q$ = 1, 1.5, 2, 11, with gas $Q$ = 1. \\

\noindent \emph{Stellar feedback:} In our first attempt, we had ignored the effect of stellar feedback in the simulation. However, we found that the stellar central surface density increased by few times of the initial value as the galaxy evolved over time. This mismatch could be adjusted by introducing stellar feedback in the model.  Simulations of galaxy formation result in galaxies with  cuspy dark matter halos  with concentrated stellar bulges \citep{Vanden2001,Barnes1996} whereas most of the LSBs are devoid of central stellar bulges. Formation of steeper stellar profiles can be stopped by removing baryons with low angular momentum. This can be driven off center by supernovae (SN) explosions \citep{Binney_2001}. The gas removal through these explosions also impart energy to dark matter particles which expand the halo near the center \citep{MoMao2004,Mashchenko2006}. This helps in maintaining a shallow dark matter profile and hence prevent the collapse of the stellar particles towards the center. \emph{Following \citet{Governato2010}, we use two parameters, star formation efficiency $\epsilon_*$ (mass ratio of stars formed to the total gas mass) as 0.05 and supernova mass fraction $\eta_{SN}$ (ratio of mass of stars formed and mass of stars lost in supernova) as 0.05 to regulate the central mass density.} This reduces the number of gas particles getting converted to star particles and sustain a shallow potential near the galaxy center. Similar values could sustain shallow dark matter halo potential of mass $10^9 M_\odot$ in dwarf galaxies . \\

\begin{figure*}
\begin{center}
\begin{tabular}{cc}
\resizebox{75mm}{!}{\includegraphics[width=5cm, height=2.7cm]{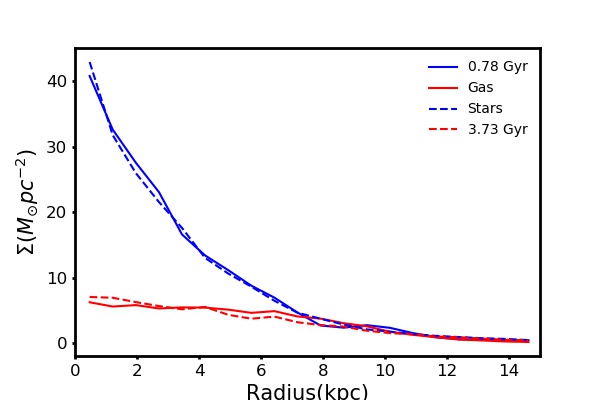}} 
\resizebox{75mm}{!} {\includegraphics[width=5cm, height=2.7cm]{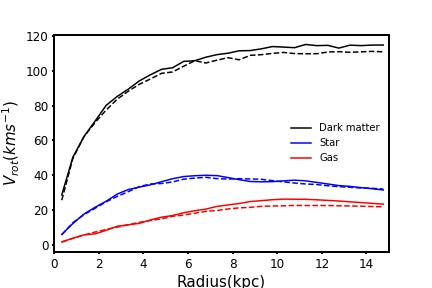}} \\
\resizebox{75mm}{!}{\includegraphics[width=3cm, height=2cm]{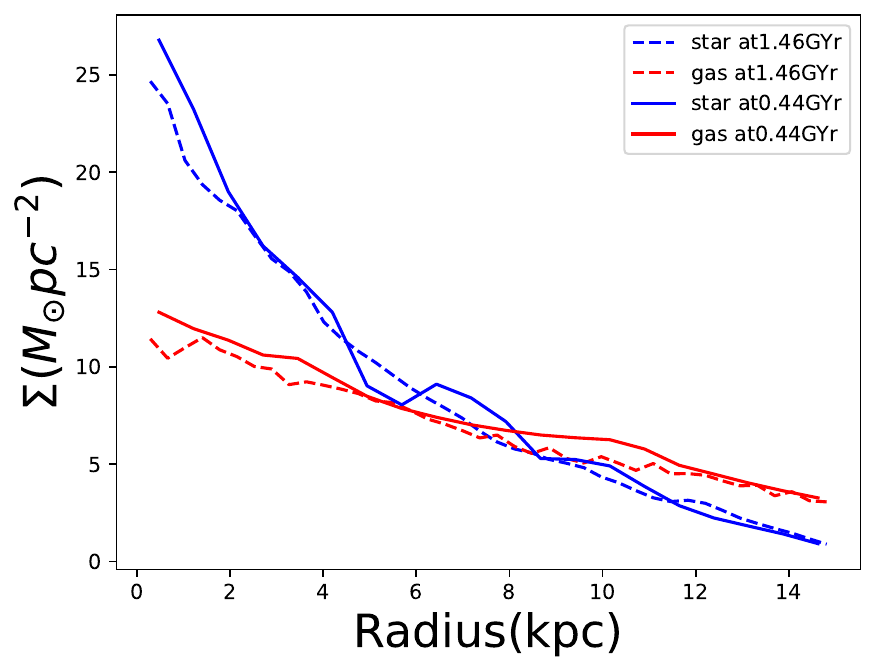}} 
\resizebox{75mm}{!}{\includegraphics[width=3cm, height=2cm]{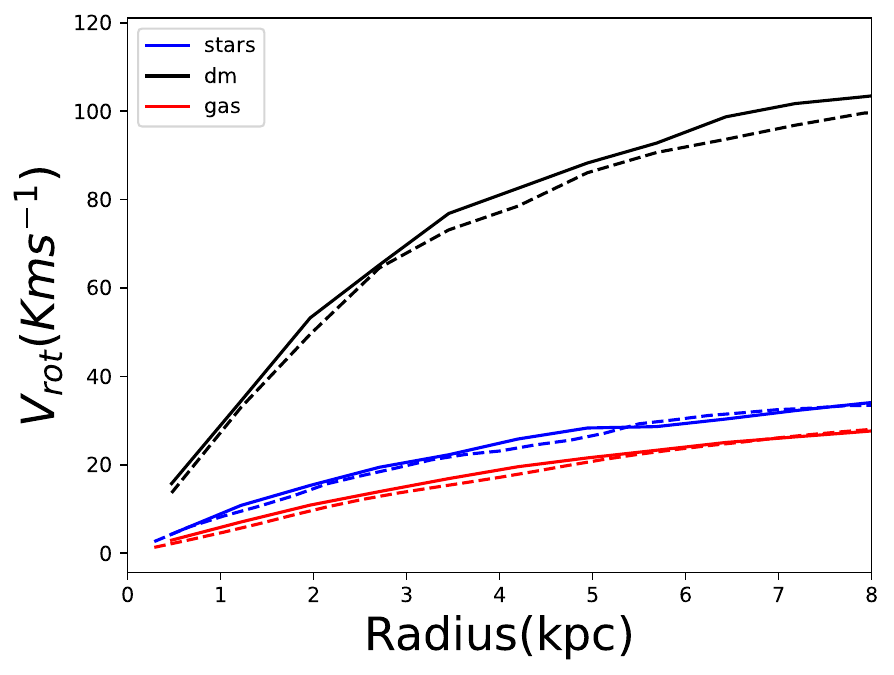}}
\end{tabular}
\end{center}
\caption{N-body + Hydrodynamical Simulations: Radial profiles of stellar and gas surface density (Left), and rotational velocity \textbf{due to} dark matter, stellar and gas potential only, respectively (Right). Results at two different epochs are shown: when the spiral just appears (solid line) and when the pitch angles of the simulated and the observed spiral are equal (dotted line). Top Panel: FGC568-VI Bottom Panel:FGC568-01}
\label{Physprops}
\end{figure*}

\noindent \emph{Observational constraints and others:} We next confirm if our simulation results comply with the constraints directly modeled from optical and HI 21cm radio-synthesis observations. \autoref{Physprops} (Top Panel) shows the radial profile of surface densities of the stellar and the gas disks \textbf{of FGC568-VI} at $0.78$ Gyr and $3.73$ Gyr, respectively in the Left Panel . The spiral pattern appears by 0.78 Gyr in the stellar disk and winds up by 3.73 Gyr. At $1.34$ Gyr, the pitch angles of the observed and the simulated spirals match the best. In the Right Panel, we plot the rotation curve again for the same epochs. We note that both the radial mass distributions, as well as the rotation curve comply with observations. We observe here that the initial conditions of the simulations were set up based on the same observational constraints, which have not evolved significantly over time. \textbf{Similarly, in the bottom panel}, we compare the rotation curves and the surface density profiles of FGC568-01 at 0.5 Gyr when it appears and at 1.4 Gyr, which is about seven dynamical times. As before, we note that there is negligible change in the profiles.
\\

\begin{figure*}
\begin{center}
\begin{tabular}{cc}
\resizebox{65mm}{!}{\includegraphics[width=5cm, height=3cm]{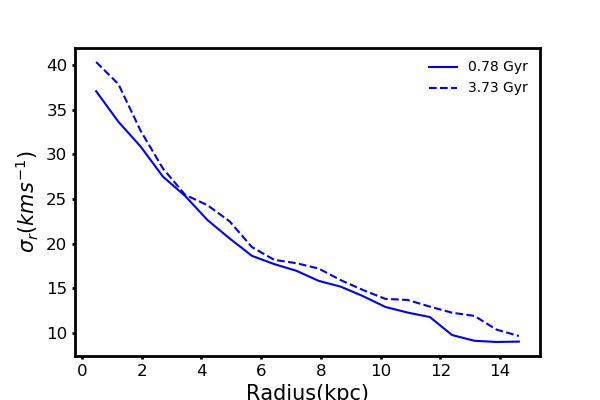}} 
\resizebox{65mm}{!}{\includegraphics[width=5cm, height=3cm]{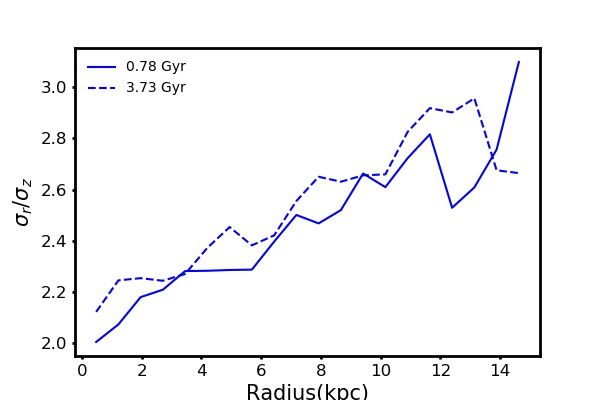}} 
\\
\resizebox{65mm}{!}{\includegraphics[width=5cm, height=3cm]{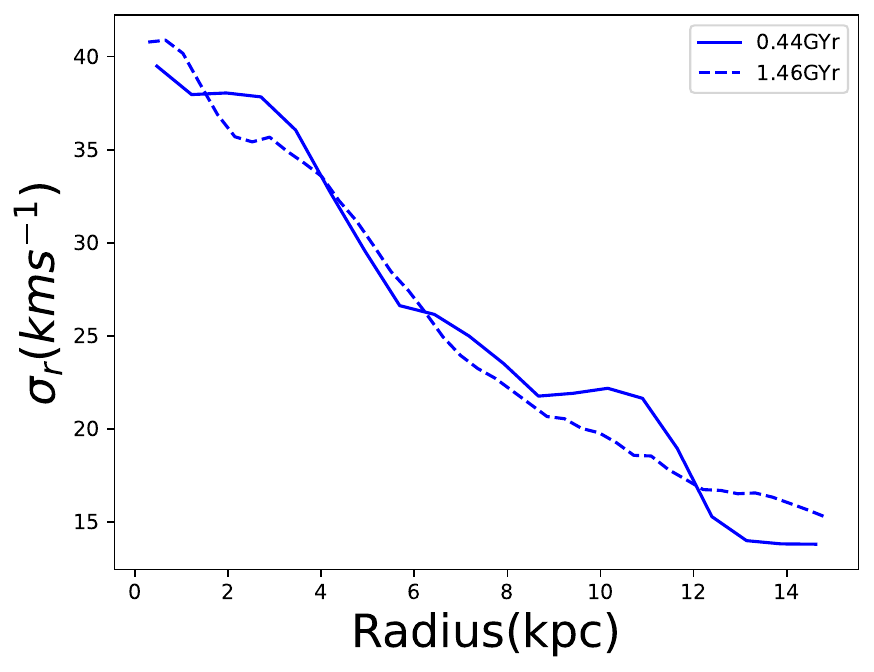}} \resizebox{65mm}{!}{\includegraphics[width=5cm, height=3cm]{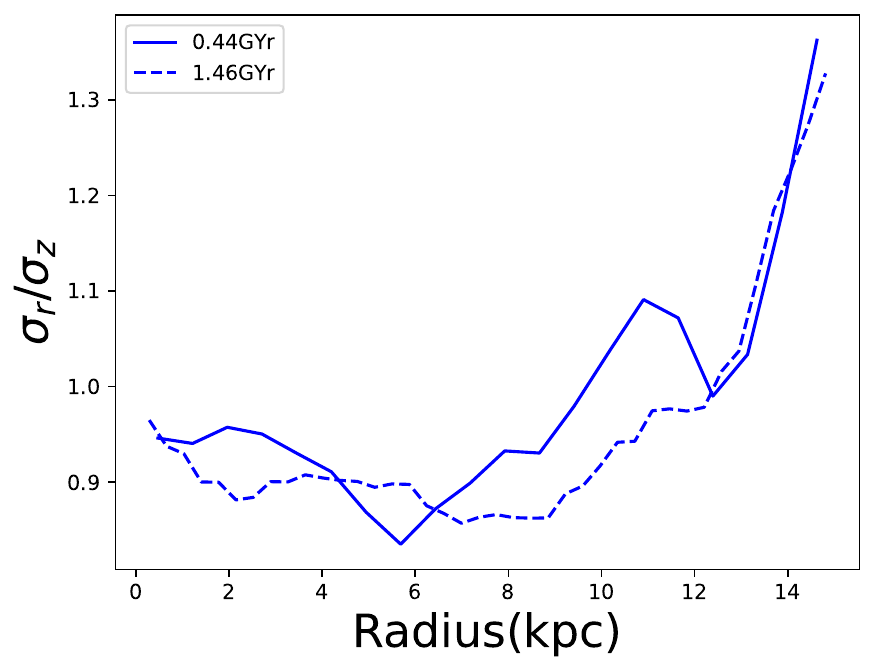}}
\end{tabular}
\end{center}
\caption{N-body + hydro-dynamical simulations: Radial profiles of the radial stellar velocity dispersion (Left Panel), and the ratio of the radial-to-vertical stellar velocity dispersion (Right Panel). Results at two different epochs are shown: when the spiral just appears (solid line) and when the pitch angles of the simulated and the observed spiral are equal (dotted line). Top Panel: FGC568-VI Bottom Panel:FGC568-01}
\label{Starprops}
\end{figure*}

\begin{figure*}
\begin{center}
\begin{tabular}{cc}
\resizebox{75mm}{!}{\includegraphics[width=5cm, height=2.7cm]{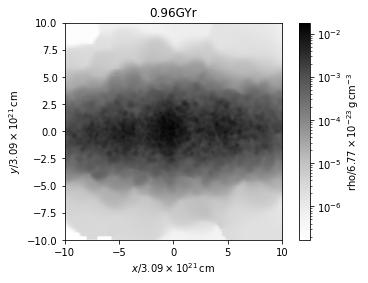}} 
\resizebox{75mm}{!} {\includegraphics[width=5cm, height=2.7cm]{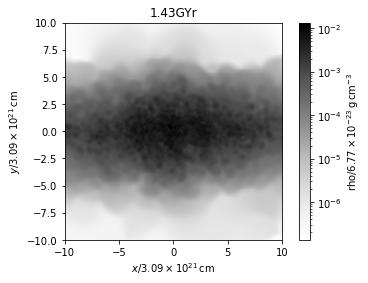}} \\
\end{tabular}
\end{center}
\caption{N-body + Hydrodynamical Simulations of FGC568-01: 
Edge-on view of the stellar disk at 0.96 Gyr (Left) and 1.43 Gyr (Right). }
\label{Bulge}
\end{figure*}

\noindent In \autoref{Starprops}, we present the radial stellar velocity dispersion (Left Panel) and the ratio of the radial-to-vertical stellar velocity dispersion (Right Panel). The top panel shows FGC568-VI and the bottom panel FGC568-01. Interestingly, for FGC568-VI, we observe $\sim 14\%$ change in both the radial stellar velocity dispersion and in the radial-to-vertical stellar velocity dispersion  $\sim R=0$  during the evolution of the galaxy, which possibly implies that disk heating is minimal due to the spiral activity, which is perhaps weak. The average value of the radial-to-vertical stellar velocity dispersion is $\sim$ 2.5, which complies with the observed fact that the radial-to-vertical velocity dispersion ratio increases from 2 to 3 for nearby late-type galaxies \citep{Gerssen2012}. For FGC568-01, however, the radial-to-vertical stellar velocity dispersion remains between 0.8 - 1 within one stellar disk scale length or so, which indicates isotropy of the stellar velocity ellipsoid. This is possibly due to the development of a small speroidal component in the galactic center, which was confirmed when the simulated galaxy was viewed edge-on. See \autoref{Bulge}. \\ 

\begin{figure*}[!h]
\begin{center}
\resizebox{65mm}{!}{\includegraphics[width=5cm, height=2.7cm]{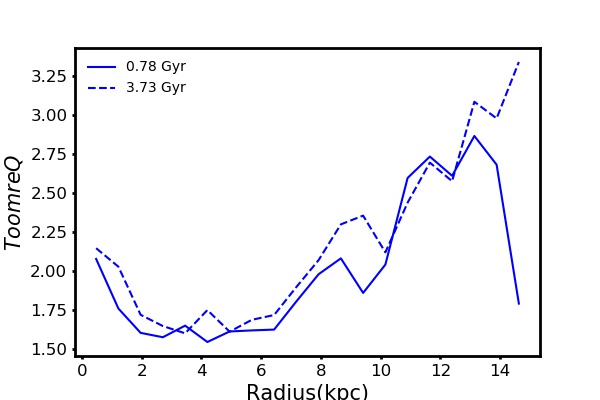}}
\resizebox{65mm}{!}{\includegraphics[width=5cm, height=2.7cm]{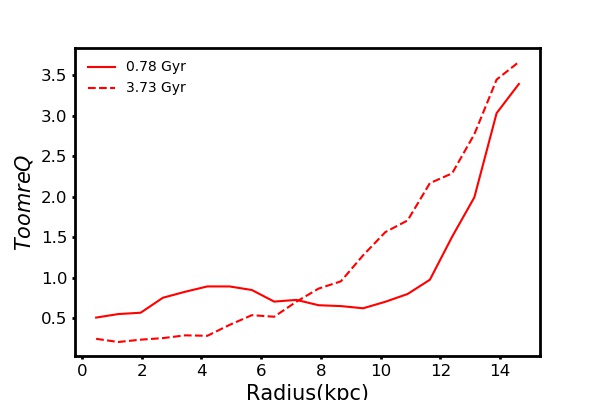}} \\
\resizebox{65mm}{!}{\includegraphics[width=4cm, height=2.7cm]{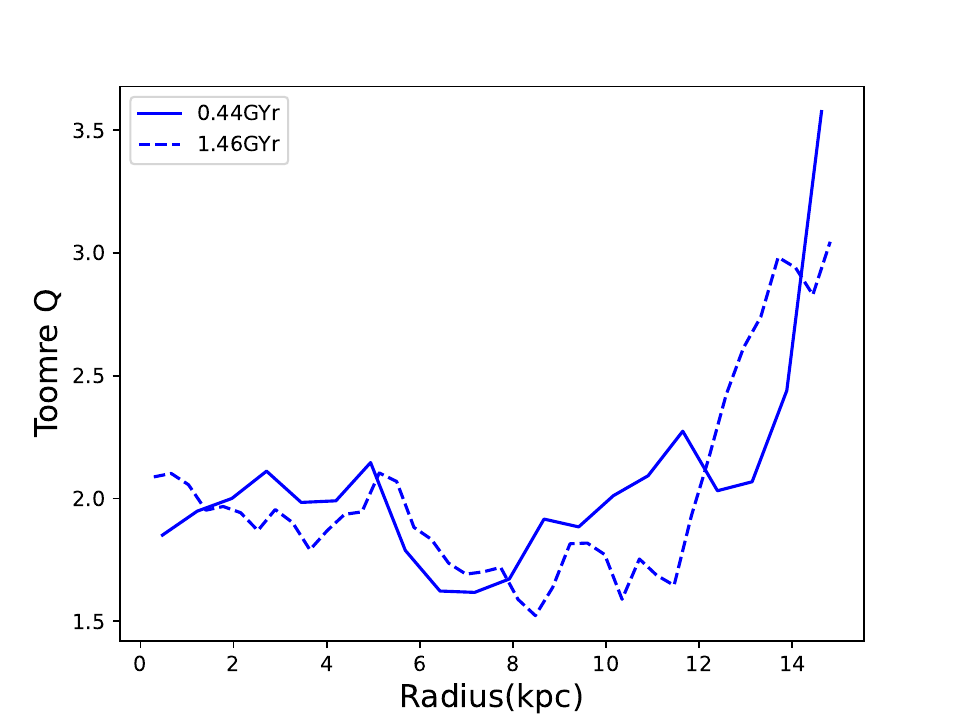}} 
\resizebox{65mm}{!}{\includegraphics[width=4cm, height=2.7cm]{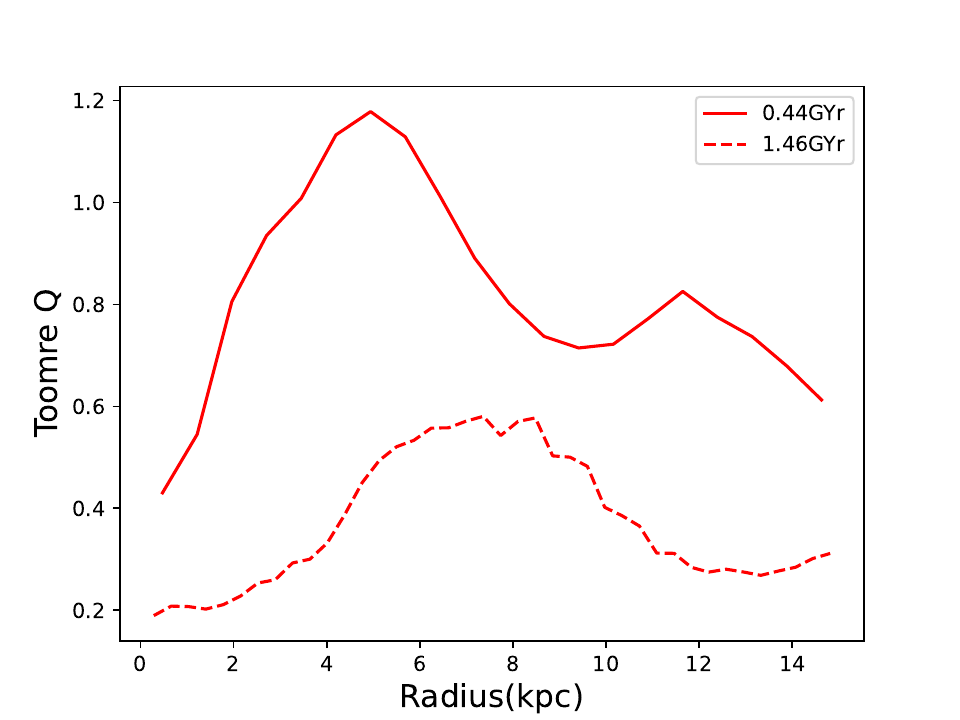}} 
\end{center}
\caption{N-body + Hydrodynamical Simulations: Top Panel: Radial profiles of Toomre Q parameter for the stellar (Left) and gas disk (Right). Results at two different epochs are shown: when the spiral just appears (solid line) and when the pitch angles of the simulated and the observed spiral are equal (dotted line). Top Panel: FGC568-VI Bottom Panel:FGC568-01.}
\label{ToomreQ}
\end{figure*}

\noindent Finally, in \autoref{ToomreQ}, top panel, we show the evolution of Toomre Q of the stellar disks of FGC568-VI (Left) and FGC 568-01 (Right). For F568-VI, the minimum stellar Toomre Q is 1.5, and increases by a factor of 3-4 as the disk evolves. For F568-01, the stellar Toomre Q remains almost unchanged on average. Similarly, for the gas disks (bottom panel), the Toomre Q increases by a factor of 5-6 in the outer disk for FGC568-VI. However, the same decreases for the gas disk of FGC568-01. \\ 

\begin{figure*}
\begin{center}
\resizebox{65mm}{!}{\includegraphics{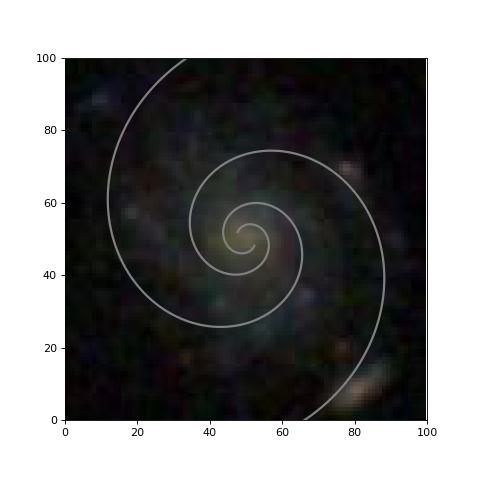}} 
\resizebox{65mm}{!} {\includegraphics{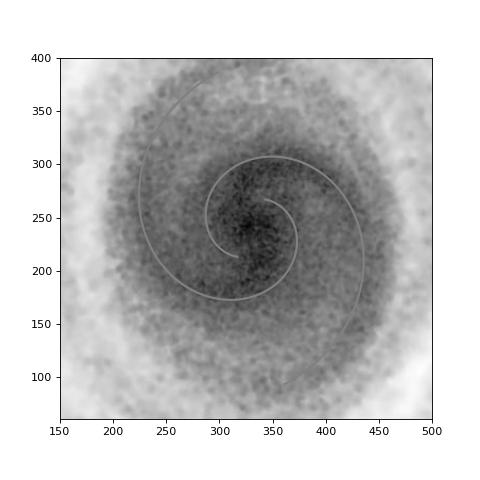}} \\
\includegraphics[scale=0.45]{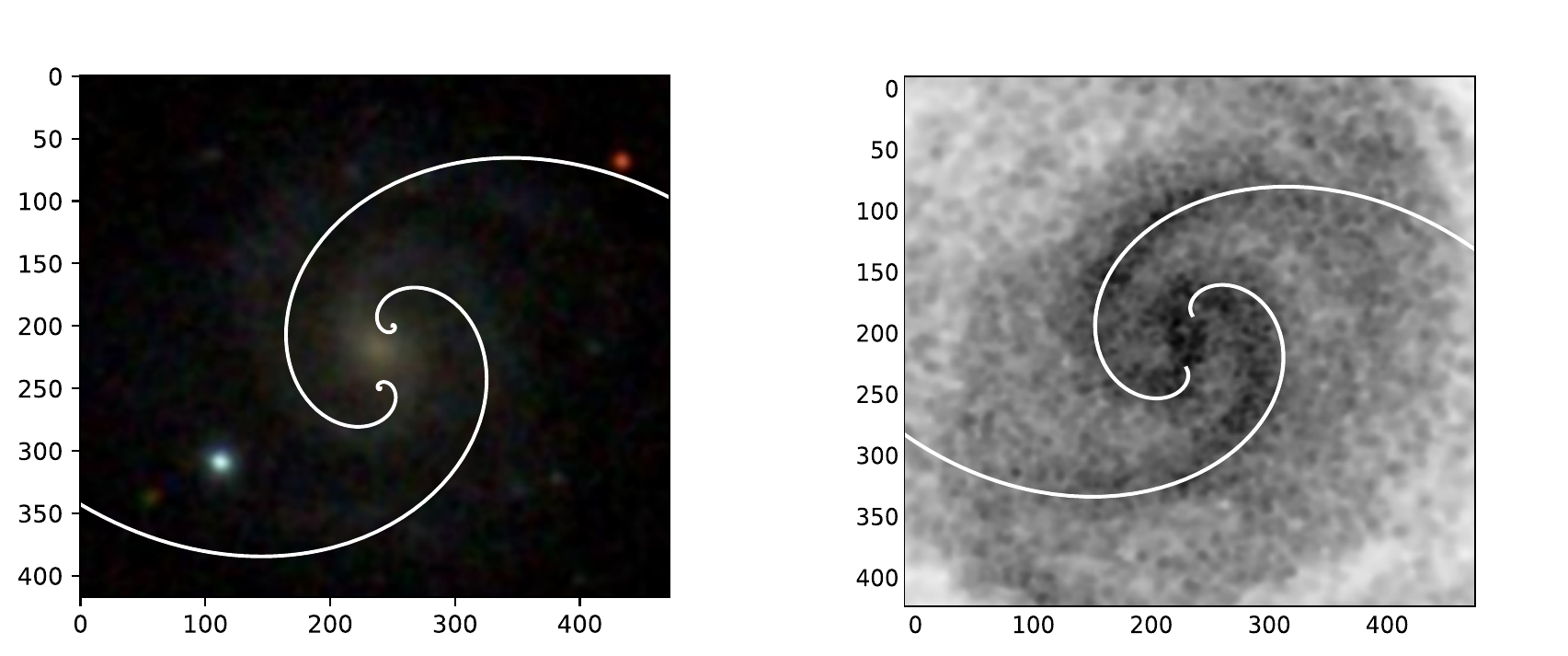} 
\end{center}
\caption{SDSS image (Left) and simulated image (Right) when the pitch angles match. In each case, a logarithmic spiral fit is superimposed. Top Panel: $F568-V1$ Bottom Panel: $F568-01$.}
\label{LGSP}
\end{figure*}

\noindent \emph{Pitch Angle:} \autoref{LGSP}, the top panel shows the observed optical image of the FGC568-VI as obtained from the SDSS \citep{SDSSDR17} (Left), and the simulated image at the epoch when the pitch angles match (Right). We have fitted a logarithmic spiral to each, which gives a pitch angle of $\mathrm{32^{\circ}}$ and $\mathrm{28^{\circ}}$ for the observed and the simulated spiral, respectively, which, therefore, mostly agree with each other. In the bottom panel, we present the same for FGC568-01. The pitch angle comes out to be $\mathrm{{29.7}^{\circ}}$ in either case. \\

\noindent \textbf{Analysis of Spiral Features} \\ 

\begin{figure*}
\begin{center}
\begin{tabular}{cc}
\resizebox{75mm}{!}{\includegraphics[width=5.5cm, height=3.5cm]{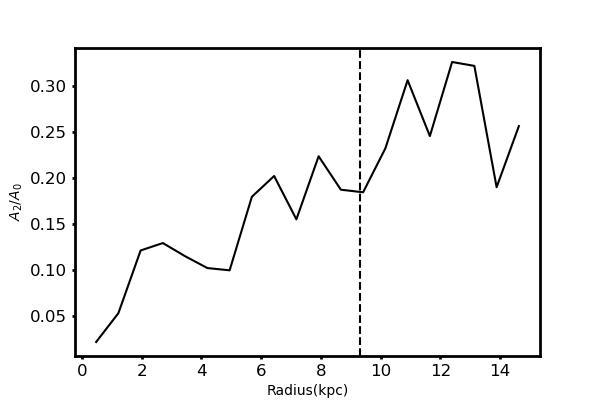}} 
\resizebox{75mm}{!}{\includegraphics[width=3.7cm, height=2.2cm]{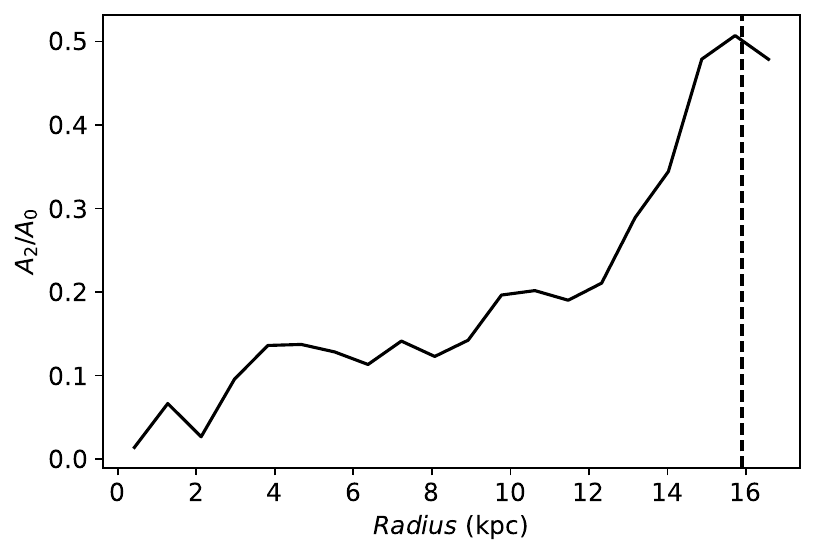}} 
\end{tabular}
\end{center}
\caption{N-body + Hydro-dynamical simulations: Ratio of the Fourier amplitudes of $\rm{m=2}$ to $\rm{m=0}$ modes from Fast Fourier Transform of FGC568-VI at 1.34 Gyr (left) and  FGC568-01 at 1.4 Gyr (right) when the pitch angles of the simulated image and observed SDSS image match with each other. Radius $\rm{3 R_d}$ is marked by dashed vertical line where $R_d$ is the exponential disc scale length.}
\label{FAmpratio}
\end{figure*}

\noindent \emph{Fast Fourier Transform} In \autoref{FAmpratio}, we present the radial profile of $\mathrm{{A_2}/{A_0}}$ i.e., the ratio of the amplitudes of the $\rm{m=2}$ to the $\rm{m=0}$ modes, obtained by performing a Fast Fourier Transform of the simulated image of the stellar disks of FGC568-VI (Left) and FGC568-01 (Right) as seen in 
\autoref{SimSnapshot1} (Top Panel, Right) and \autoref{SimSnapshot2} (Bottom Panel, Right), respectively. In both cases, the amplitude of the spiral $\mathrm{A_2/A_0 > 0.2}$  is sufficient to identify it as a spiral structure in simulations \citep{Dobbs_2014}. We note that the value of the ratio always lies well below $\rm{0.4}$, thus indicating that the spiral arm formed is weak. \\

\begin{figure*}
\begin{center}
\begin{tabular}{cccc}
\resizebox{75mm}{!}{\includegraphics{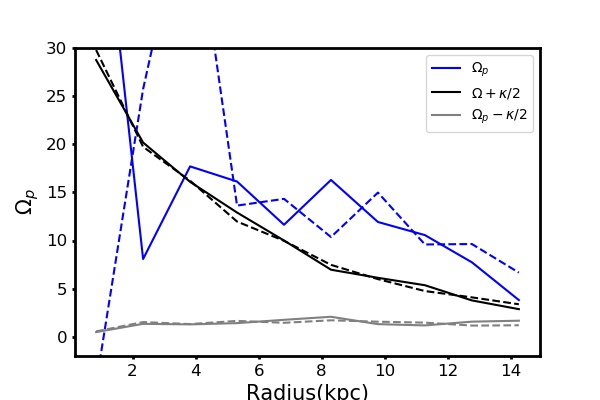}} 
\resizebox{65mm}{!}{\includegraphics{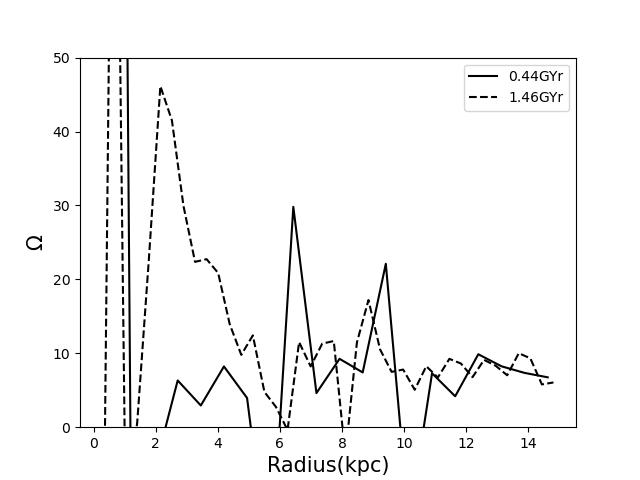}} 
\end{tabular}
\end{center}

\caption{ N-body + Hydrodynamical Simulations:  Radial variation of pattern speed of the $\rm{m=2}$ mode in the stellar disk of (i) FGC568-VI at 0.78 Gyr (solid line) and 1.34 Gyr (dotted line). (Left Panel) (ii) FGC568-01 at 0.96 Gyr (solid line) and 1.4 Gyr (dotted line)(Right Panel).}
\label{PatternSpd}
\end{figure*}
\begin{figure*}
\begin{center}
\begin{tabular}{cc}
\resizebox{55mm}{!}{\includegraphics{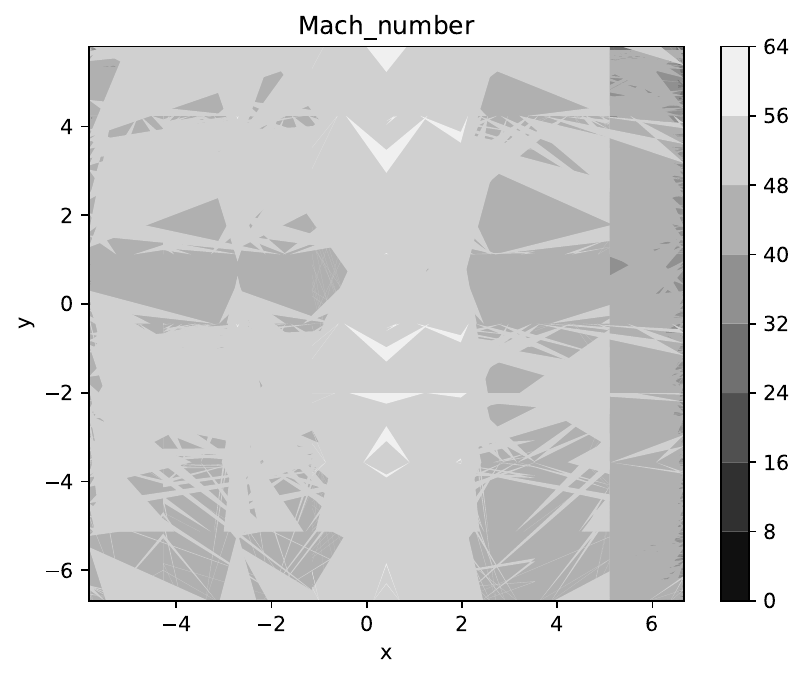}} 
\resizebox{63mm}{!}{\includegraphics{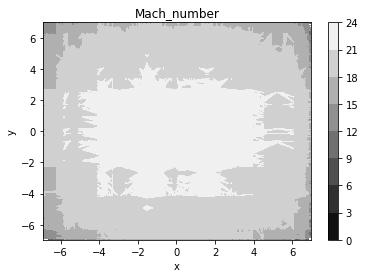}} 
\end{tabular}
\end{center}
\caption{Mach no in the X-Y plane of the gas disk in LSB F568-V1[left] and LSB F568-01[right]}
\label{MachNo}
\end{figure*}

\noindent In \autoref{PatternSpd}, (Left Panel), we present the radial profile of the pattern speed of the $m=2$ mode of FGC568-VI at two epochs: when the spiral features just appear (solid line) and when the pitch-angles of the observed and the simulated images are equal (dotted line). For FGC568-VI, we note that the pattern speed fluctuates quite a lot within $2 R_d$ beyond which it slowly falls off with radius. Also, beyond $2 R_d$, the pattern speed remains almost constant with time, the average value being 15 kms$^{-1}$. Interestingly, the pattern speed obtained from our simulation using 2D-FFT lies close to the pattern speed for the most unstable mode in the global mode analysis. In \autoref{PatternSpd} (Right Panel), we present the pattern speed of the spiral for FGC568-01. Similar to FGC568-VI, no well-defined pattern speed exists. The average value of the pattern speed is about 10 kms$^{-1}$. The values of the pitch angles and the pattern speeds are presented in \autoref{ResRAMSES}. \\

\begin{table}
	\centering
	\caption{Results from RAMSES simulation}
	\label{ResRAMSES}
	\begin{tabular}{lcc} 
        \hline 
        &F568-VI&F568-01\\
		\hline
		\underline{Pitch angle} \\
		Observed (SDSS image) & $\mathrm{32^{\circ}}$&$\mathrm{29.7^{\circ}}$   \\
		Simulated (1.34 Gyr) & $\mathrm{28^{\circ}}$& $\mathrm{29.7^{\circ}}$  \\
		\hline
		\underline{Pattern speed} & $\mathrm{km s^{-1} kpc^{-1}}$\\
	    2D-FFT & 15&10 \\
		\hline
	\end{tabular}
\end{table}

\noindent \emph{A stationary density wave or a transient spiral?} The primary signature of a stationary, density wave is a pattern speed constant with radius and time. In the case of F568-VI, we note that the pattern speed of the $m=2$ mode survives for 2.1 Gyrs, and the average pattern speed varies between 14 - 16  $\rm{kms^{-1}{kpc}^{-1}}$; the shearing rate at each epoch being $\sim$ 1  $\rm{kms^{-1}{kpc}^{-2}}$. The radial dependence of the pattern speed has been observed, for instance, in M51 \citep{Meidt2005}. Further, the necessary condition for the existence of a stationary density wave is the presence of the Inner and Outer Lindblad resonances. During the nascent stages of the formation of the spiral, when linear analysis was still applicable, we checked for the presence of Inner and Outer Lindblad resonances. However, their presence was not quite apparent (See \autoref{PatternSpd}, Left Panel). The above observations, taken together, constitute the fingerprint of a transient spiral feature. The argument of the superposition of modes leading to transient spirals in simulations is not applicable in this case, as we specifically picked up the $m=2$ mode. Thus we conclude that the spiral features observed in FGC568 VI constitute a \emph{transient spiral pattern} driven by the quadrupolar potential of an oblate dark matter halo. Similarly, we can argue that the spiral in FGC568-01 is not a stationary density wave as its pattern speed is toy constant with radius, and keeps fluctuating with time.\\
\\

\noindent \emph{Why does the gas disk not form spiral arms?} Both stars and gas have been considered as self-gravitating components in our simulations. But, unlike the stars, gas, constitutes a collisional medium, and hence, in general, will not form a self-gravitating pattern like the spiral arm, due to the excitation of shock waves (See, for example, \cite{combes1995}). In \autoref{MachNo}, we have now presented the map of the Mach Number of the gas disks of FGC568-VI (Left) and FGC568-01 (Right).  
Interestingly, between F568-VI and F568-01, the former has a gas surface density few times higher than the latter. In addition, F568-VI has a much higher mach number ($\sim$ 40 - 55) than F568-01($\sim$ 20 - 24). The combined effect of the self-gravity and shock waves is reflected in the disks of the two LSBs. As expected, very weak structures are seen in the gas disk of F568-01 due to its larger self-gravity and lower mach number, but no such features are noticeable at all in F568-VI.

\section{conclusion}

\noindent We model the spiral features of the proto-typical LSBs FGC568-VI and FGC568-01 using analytical methods and numerical simulations. If we consider the LSB disk to be a 2-component system of gravitationally-coupled stars and gas hosted in a spherical dark matter halo, we find that it is stable against local, non-axisymmetric perturbations and hence unlikely to produce local spiral features. However, the LSB disk assumed to be a single component system of self-gravitating stars, also subjected to the external potential of the gas disc and a spherical dark matter halo is found to be unstable to a global, spiral instability. Finally, using N-body + hydrodynamical simulations using the publicly-available code RAMSES, we note that the observed spiral features of the LSB can be modeled as a transient, global spiral pattern existing for 2.1 Gyrs, with an average pattern speed of 14 - 16  $\rm{kms^{-1}{kpc}^{-1}}$ in FGC568-VI. In FGC568-01, the same exists beyond 3 Gyrs, with a pattern speed of 10  $\rm{kms^{-1}{kpc}^{-1}}$. The spiral features are driven by the quadrupolar potential of an oblate dark matter halo with a vertical-to-planar axes ratio of 0.7 and a spin parameter of 0.02.

\bibliography{Simulation_new}
\bibliographystyle{aasjournal}
\end{document}